\documentstyle[preprint,eqsecnum,prd,aps,epsf]{revtex}
\begin{document} 
\tightenlines
\preprint{
\vbox{
\halign{&##\hfil\cr
	& ANL-HEP-PR-97-21 \cr
        & UFIFT-HEP-97-12  \cr
	& JLAB-THY-97-40 \cr}}}

\title{
Rapidity Correlations and $\Delta G $ from Prompt Photon plus 
Jet Production in Polarized $pp$ Collisions}

\author{
Sanghyeon Chang$^{1}$\footnote{
        E-mail address: schang@phys.ufl.edu},
        Claudio Corian\`{o}$^{2}$\footnote{
        E-mail address: coriano@jlabs2.jlab.org}   
        $and$\, L. E. Gordon$^{3}$ \footnote 
{E-mail address: gordon@hep.anl.gov}   }
\address{ $^{1}$   Institute for Fundamental Theory, Phys. Dept., 
        Univ. of Florida, Gainesville, FL 32611\\
$^{2}$  Theory Group, Jefferson Lab, Newport News, VA 23606 \\
$^{3}$ High Energy Physics Division, Argonne National Laboratory,
	Argonne, IL 60439, USA }
\maketitle
\begin{abstract} 
A study of prompt photon plus associated jet production is performed
at next-to-leading order (O($\alpha\alpha_s^2$)) in QCD at $\sqrt{S}=200-500$ 
GeV, appropriate for the RHIC polarized $\vec{p}\vec{p}$ collider experiment.
Momentum correlations between the jet and photon are examined and the
utility of the process as a method for constraining the size and shape of the 
polarized gluon density of the proton $\Delta G$ is examined. 
\end{abstract}
\vspace{0.2in}


\setcounter{footnote}{0}

\def\beq{\begin{equation}}
\def\eeq{\end{equation}}
\def\beqn{\begin{eqnarray}}
\def\eeqn{\end{eqnarray}}

\def\ie{{\it i.e.}}
\def\eg{{\it e.g.}}
\def\half{{\textstyle{1\over 2}}}
\def\third{{\textstyle {1\over3}}}
\def\quarter{{\textstyle {1\over4}}}
\def\m{{\tt -}}

\def\p{{\tt +}}

\def\slash#1{#1\hskip-6pt/\hskip6pt}
\def\slk{\slash{k}}
\def\GeV{\,{\rm GeV}}
\def\TeV{\,{\rm TeV}}
\def\y{\,{\rm y}}

\def\l{\langle}
\def\r{\rangle}

\setcounter{footnote}{0}
\newcommand{\beqa}{\begin{eqnarray}}
\newcommand{\eeqa}{\end{eqnarray}}
\newcommand{\eps}{\epsilon}

\section{Introduction }

With the advent of RHIC, the Relativistic Heavy Ion Collider at Brookhaven, 
QCD will enter into a new interesting phase in which polarized high energy 
collisions will become a standard tool of analysis in high energy physics. 
RHIC, as a $pp$ collider, will be endowed with a very high luminosity 
(500 pb$^{-1}$), not easily accessible to $p\bar{p}$ colliders and will be 
spanning a center of mass energy range between 50 to 500 GeV. 

One of the main programs at RHIC will be to nail down the size and the 
shape of the polarized parton distributions which, at the moment, suffer from 
significant model dependence, especially in the gluon contribution
($\Delta G$). Interest in spin physics has grown in the last few years 
thanks to the various polarized DIS experiments and to the existing proposal 
for the construction of a $pp$ collider at DESY (HERA-$\vec{N}$) running at 
lower energy ($\approx 50$ GeV).  

More generally, besides the important information on the spin structure of the 
nucleon which can be gathered from these experiments, it is obvious that one 
might try to look for physics beyond the Standard Model using polarization as 
a tool to suppress unwanted background and to enhance specific signals. The 
studies of chiral couplings in the Standard Model or even compositeness will 
require precise measurements of the hadronic background. There have been 
many attempts in the last few years at creating a 
backbone-program for RHIC by analyzing a set of processes in leading order and 
by defining suitable observables which are more easily accessible to the 
experimental investigations. 

More recently, next to leading order (NLO) studies of various processes 
(see for instance \cite{CG}) have 
been presented. A partial review of these developments can be found in 
\cite{cheng}. Among the most interesting processes which require an accurate 
theoretical determination are single and double prompt photon production, 
single and 2-jet production and the Drell-Yan lepton-pair production. 
Work on the Drell Yan cross section at NLO has been presented 
\cite{CCFG,CCE}, limited at the moment to the non-singlet sector, which is sensitive to the polarization of the quark distributions. Due to the fact 
that the initial partons are longitudinally (or transversely) polarized, the 
calculation of the hard scatterings are far more involved than in the 
unpolarized case. 

Although the studies of the $total$ cross sections for some of these and other 
related processes will be necessary in order to guarantee experimental 
observability of the cross sections and thus the spin asymmetries, which are 
usually 
estimated to be small, the study of the event-structure of these processes can 
provide interesting new insights into the spin distributions. In this paper we 
study and compare the structure of prompt photon plus single jet production in 
the polarized and unpolarized cases at RHIC center-of-mass energies.

\section{ Monte Carlo calculations at NLO}

Compared to total cross sections, distributions usually reveal more
detail about the underlying hard scattering mechanisms as well as more
information on the $x$-dependence of the model structure functions. The
drawback is that they are generally more difficult to define theoretically to 
NLO, since many of them are affected by non-canceling infrared divergences. 
Modern developments in combined Analytic/Monte Carlo techniques allow to get 
exact numerical results for the NLO corrections in a reasonable amount of time.
In addition they allow great flexibility in placing experimental
selections such as jet definitions and photon isolation cuts on the
cross sections, thereby allowing a more realistic and direct comparison with 
data. 
   
The analytical part of the calculation involves the 1) exact (analytical) 
evaluation of the virtual corrections and 2) the exposure (by a cutoff 
regularization) of all the mass singularities in the real emissions. The
rest of the phase space is then integrated over numerically.
We omit a general presentation of the method which can be found 
elsewhere and focus on the study at NLO of the rapidity correlations in photon 
plus associated jet production. 

In the next sections, after a brief overview of the various contributions to 
the cross section, we move to a study of the rapidity correlations between the 
photon and the jet. The analysis presented is similar in spirit to the study 
of momentum correlation given in ref . \cite{berger,bailey}, with the due 
modifications.

\section{The Photon plus Jet Cross Section}

The main goal at polarized hadron colliders is the study of the 
polarized parton distributions of the nucleon which are defined by
\begin{equation}
\Delta f_i(x,M^2)=f_i^+(x,M^2)-f^-_i(x,M^2).
\end{equation}
The corresponding unpolarized distributions are defined by
\begin{equation}
f_i(x,M^2)=f_i^+(x,M^2)+f^-_i(x,M^2),
\end{equation}
where
 $f^+_i$ and $f^-_i$ represent the distribution of partons of
type $i$ with positive and negative helicities respectively, with respect to 
that of the parent hadron. The hard subprocess scattering cross sections are 
defined by 
\begin{equation}
\Delta\hat{\sigma}_{ij}=\frac{1}{2}\left( \sigma^{++}-\sigma^{+-}\right)
\end{equation}
and
\begin{equation}
\hat{\sigma}_{ij}=\frac{1}{2}\left( \sigma^{++}+\sigma^{+-}\right).
\end{equation}
for the polarized and unpolarized cases respectively.

It has been observed by many authors that the cross section for prompt photon 
production is dominated by the subprocess $qg\rightarrow\gamma q$ in hadronic 
collisions already in leading order. This means that the cross section,
if properly understood, could potentially prove very useful for providing 
information on the unpolarized gluon densities, $g(x,Q^2)$, of 
hadrons. Original Born level studies of this cross section also
indicated that the same is true in the polarized case. It has therefore been 
suggested that it may prove useful in pinning down the polarized gluon 
densities \cite{bergerqiu}. In this context it has been examined quite 
extensively in leading, and more recently in next-to-leading order 
\cite{contogouris} \cite{gorvogel}. The most recent NLO study
\cite{gordon1} included the effects of photon isolation on the cross
section. The results of all the NLO studies confirmed the conclusions from 
the LO ones that the cross section is very sensitive to $\Delta G$ and
that the asymmetries are perturbatively stable.

Recently the photon plus jet cross section was studied in LO
\cite{nowak} and NLO \cite{gordon2} and HERA-$\vec{\rm N}$ cms energies.
The main conclusions from these studies is that this cross section will
give more detail about the $x$-shape of the polarized gluon
distribution than the single prompt photon cross section. The present
calculation follows along the same lines as that in ref.\cite{gordon2}
and we therefore do not give more detail about the contributing
subprocesses here, except to say that the fragmentation contributions
are estimated in LO here as well. In the present case, before isolation,
these contributions are numerically more important due to the higher cms
energy of the RHIC collider. The results of ref.\cite{gordon1} also
showed that isolation significantly reduced these contributions and that
that their presence, although affecting the predictions for the cross
sections, did not affect the asymmetries very much.    
  
\subsection{Rapidity Correlations to LO and NLO}

As mentioned above, although single inclusive prompt photon production will 
definitely be very  important for constraining the size of $\Delta G$, 
information on the detailed $x$-shape of the distribution will not 
be as easily extracted. This is because the calculation of the inclusive
cross section involves one convolution over the momentum fractions, $x$, of 
the initial partons. In fact, at hadron level, the factorized hadronic total 
cross section is  generically denoted as
\beqa
\Delta\sigma=\sum_{i,j}\int_{0}^{1} dx_1 dx_2 \Delta f_i(x_1,) \Delta f_j(x_2) \int d\Delta \sigma.
\label{one}
\eeqa
where we sum over all the partons $i,j$. 
The practical effect of the integral over the $x_i$ is that a measurement of
the kinematic variables of the photon is not sufficient to determine
them. If, on the other hand, one or more of the jets
produced in the reaction is also tagged, no convolution is involved in
the calculation and the cross section is directly proportional to the
parton densities. 

The cross section of interest here is the triple differential cross
section 
\begin{displaymath}
\frac{d^3\Delta\sigma^{\gamma J}}{dp_T^\gamma d\eta^\gamma d\eta^J},
\end{displaymath}
where $\eta^\gamma$ and $\eta^J$ are the pseudorapidities of the photon
and jet respectively and  $p_T^\gamma$ is the transverse momentum of the
photon.

We use light-cone coordinates 
$$ p_1=p_1^+ n^+= x_1 Q n^+\hspace{1cm}p_2=p_2^-n^-=x_2 Q n^- 
\hspace{1 cm} n^\pm= 1/ \sqrt{2}( 1, 
\bf{0\perp}, \pm 1) $$
with $Q=\sqrt{S}/2$ denoting the $\pm$ components on the light cone of the 
two incoming hadrons of momenta $P_1$ and $P_2$. We also set $p_1=x_1 P_1$ and 
$p_2= x_2 P_2$ for the two partons that enter the hard scattering. 

The 4-dimensional $\delta$ function and the integration 
variables $k_i$ are also rewritten on the light cone 
$(k_i=k_i^+,k_i^- k_{i\perp})$
 and after simple manipulations we get

\beq
\Delta \sigma \sim\sum_{i j}\int_{0}^1 dx_1 \int_0^1 dx_2 \Delta f_i(x_1)\Delta f_j(x_2)
\int dk_1^+ dk_2^- d^2 k_{1\perp}\delta(k_1^2)\delta(k_2^2)\Delta |M|^2.
\eeq

We have set  
\beq
k_1^+= {k_\perp\over \sqrt{2}}e^{y_1}  \hspace{1cm}
k_2^-={k_\perp\over \sqrt{2}}e^{-y_2}.
\eeq

We have defined the two rapidities 

\beq
y_i ={1\over 2}\log {k_i^+\over k_i^-} \hspace{1cm} i=1,2 
\eeq
and introduced the rapidity difference $\Delta y=y_1-y_2$. 

We easily get 
\beq
\frac{d^3\Delta\sigma^{\gamma J}}{dp_T^\gamma d\eta^\gamma
d\eta^J}\equiv
{d\Delta\sigma \over d k_\perp d y_1 d y_2}= 2 k_\perp \overline{x_1}
\overline{x_2}
\sum_{i j}\Delta f_i(\overline{x_1})\Delta f_j(\overline{x_2}) 
{d\Delta \sigma\over d t}
\label{mainq}
\eeq
A derivation of this result is illustrated in the appendix. 
A similar result applies to the unpolarized case, except that the
polarized cross sections and structure functions are replaced by the
corresponding unpolarized ones. In this next-to-leading order
calculation a jet definition is required. Throughout we use the Snowmass
\cite{snowmass} jet definition.

All the 2-to-2 contributions (Born and virtual)  to the photon-plus-jet 
cross section give contributions with structure functions sampled at 
fixed kinematical points $x_i$. 
Thus the double longitudinal asymmetry $A_{LL}$, defined as the ratio of
the polarized to the unpolarized cross section,
is directly proportional to the ratio $\Delta G/g$ \cite{nowak} in kinematic 
regions where other subprocesses such as $q\bar{q}$ scattering can be neglected.
This guarantees sensitivity of the asymmetries to $\Delta G$.

\section{Results}

All results are displayed for $\vec{p}\vec{p}$ collisions at the center-of-mass
energy $\sqrt{s}=200$ and $500$ GeV which are energies typical for the
RHIC experiment at Brookhaven. For the unpolarized cross section the
CTEQ4M parton densities \cite{CTEQ} are used throughout, and the value of
$\Lambda_{\overline{MS}}$ corresponding to this distribution is also used.
Use of other unpolarized parton densities at the $x$-values probed
here do not yield significantly different results.
For the polarized case the GRSV \cite{grsv} and GS \cite{gs}
distributions are used with the corresponding values for 
$\Lambda_{\overline{MS}}$. The authors of ref.\cite{grsv} and \cite{gs} have 
proposed various parameterizations of the polarized parton densities
differing mainly in the choice of input for the polarized gluon density
$\Delta G$. In the case of the GRSV distributions we use the `valence'
set which corresponds to a fit of the available DIS data (referred to by
the authors as the `fitted' $\Delta G$ scenario), the large
gluon fit which assumes that $\Delta G(x,Q_0^2)=g(x,Q_0^2)$ at input (the
'$\Delta G=g$' scenario) and 
the small gluon fit which uses $\Delta G(x,Q_0^2)=0$ at the input scale
(the '$\Delta G=0$' scenario), which in this case starts at the very low
value of $Q_0^2=0.34$ GeV$^2$. The latter two distributions are intended
to represent extreme choices for $\Delta G$. 
These parameterizations give gluon densities which differ in their absolute
sizes as well as in their $x$-shape. The GS parameterizations provide
three fits to the data; GS A, GS B and GS C. It has been shown that the
GS A and GS B distributions do not differ very much from the $\Delta
G=g$ and fitted $\Delta G$ sets of GRSV respectively, whereas the the GS C 
set is widely
different from any of the others. We shall present
distributions using the three GRSV sets discussed above, along with the
GS C set for comparison. For the fragmentation functions we use
the LO asymptotic parameterization of ref.\cite{owens}. As will be shown,
the choice of fragmentation functions makes very little difference to
the predictions, since these processes account for only a small fraction
of the cross section.

The renormalization, factorization, and  fragmentation scales are set to a 
common value $\mu = p_T^{\gamma}$ unless otherwise stated.  
Since there are two `particles' in the final state,
the jet and the photon, both of whose transverse momenta are
large, an alternative choice might be $\mu = p_T^J$ or some function
of $p_T^{\gamma}$ and $p_T^J$.  The results of the calculations
show that the magnitudes of $p_T^{\gamma}$ and $p_T^J$ tend to be 
comparable and that dependence of the asymmetries on $\mu$ is slight,
although the individual cross sections may vary significantly with $\mu$.  
Therefore, choices of $\mu$ different from $\mu = p_T^{\gamma}$ should not 
produce significantly different predictions for the asymmetries . The two loop 
expression for
$\alpha_s(Q^2)$ is used throughout, with the number of flavors fixed at
$N_f=4$, although the contribution from charm was verified to be negligible
at the energies considered and is not included. None of the currently 
available polarized distributions include a parameterization of the charm
quark distribution. A new NLO parameterization is in preparation which
includes charm, but is not yet available for this study
\cite{ramsey}. Finally, the values of both the jet cone size and
photon isolation cones are fixed at 
$R_J=0.7$ and $r_{\gamma}=0.7$ respectively, unless otherwise stated.

Figs.1a and b show the triple differential cross section as a function of 
$p^\gamma_T$ of the photon for the various parameterizations at
$\sqrt{S}=500$ and $200$ GeV respectively. The unpolarized 
cross sections are shown for comparison. The curves were obtained 
by averaging over bins $\Delta p_T^\gamma=1$ GeV and the 
restriction $p_T^J\geq 10$ GeV was imposed. In addition both the photon
and jet rapidities are averaged over the central region, $-0.5\leq
\eta^{\gamma},\eta^J\leq 0.5$. All the parameterizations give distributions 
which are distinctly different in both their shapes and sizes. Their
relative sizes are in direct relation to the sizes and shapes of their 
respective gluon distributions.  This is most obvious for the GS C 
parameterization which has the most distinct gluon distribution, being negative
over part of the $x$-range. The curves show a rise between $p_T^\gamma=10$
and 11 GeV because of the restriction $p_T^J\geq 10$ GeV and the fact
that the bin centered at $10$ GeV is averaged over the range $9.5\leq
p_T^{\gamma}\geq 10.5$ GeV. Above $p_T^\gamma=10$ GeV both the two- and
three-body contributions to the cross section are finite whereas below
this value, only the latter are finite since the two-body contributions
always produce a photon balancing the $p_T$ of the jet.

Figs.1c and 1d show the asymmetries for 1a and 1b respectively. There
are clear distinctions between the predictions for the various
parameterizations which will certainly make them distinguishable in the
experiments. As expected, at the lower $\sqrt{S}$ the asymmetries are
larger, but the corresponding cross sections are smaller. Combining both
sets of results will nevertheless suggests that a measurement of the polarized
gluon distribution will be possible between $x=0.04$ and $x=0.5$. 

In Figs.2a and 2b we look at distributions in $\eta^\gamma$ at $p_T^\gamma=10$
GeV with the restriction $p_T^J \geq 10$ GeV still imposed. In both
cases $\sqrt{S}=200$ GeV. In Fig.2a the jet 
is restricted to be in the central rapidity region, $-0.5\leq \eta^J \leq 0.5$,
and in  Fig.2c it is restricted to the forward region, $0.5\leq
\eta^J\leq 1.5$
. The textures of the different curves follow the same conventions as in
figs.1a-d. A visual comparison of the figures show clearly that when the
jet is in the central rapidity region the, the photon rapidity
distribution peaks at $\eta^{\gamma}=0$, whereas when the jet is
restricted to the forward rapidity region, the $\eta^{\gamma}$ also peaks
in this region. This positive rapidity correlation between the photon
and jet is present for both the polarized and unpolarized
cross sections although it is clearly stronger in the unpolarized case. 

Fig.2c which shows the corresponding asymmetries for the
curves in fig.2a, verifies that the distributions are symmetric and also
shows that they have distinctive shapes. The corresponding asymmetry
curves for fig.2b all rise sharply in the negative rapidity region,
i.e., in the opposite direction to the region where the rapidity
distributions peak. Thus the asymmetries display a negative rapidity
correlation between the photon and jet. An explanation for this
behavior was given in ref.\cite{gordon2} in terms of competing effects
between the subprocess matrix elements, which tend generate positive
rapidity correlations, and the polarized parton distributions which tend
to produce negative correlations.    

In figs.3a and 3b rapidity distributions are plotted for the jet using
similar cuts to those used in figs.2a and 2b. In fig.3a, the photon is
restricted to be in the central rapidity region whereas in 3b it is
restricted to be in the forward region. As before, there is a net positive
rapidity correlation between the photon and jet for the polarized as
well as the unpolarized cases, but as fig.3d shows, the effect is
stronger for the unpolarized case. This leads the asymmetries to peak at
negative rapidities of the jet.    

One of the striking features of the asymmetry curves in figs.3c and 3d
is the differences in their shapes as compared to each other and as
compared to those of figs.2c and 2d. The parameterizations with the larger
polarized gluon distributions the ``fitted" $\Delta G$ and $\Delta G = g$
scenarios give asymmetries which decrease as $\eta^J$ moves away from
the central region in fig.3c and away from $\eta^J \sim -1.5$ in fig.3d. 
The differences between these curves and those of fig.2 are explained by 
the asymmetric $p_T$ cuts between the photon and jet which affects the 
phase space available for jet or prompt photon production differently.

The differences between the shapes of the asymmetries for the various models 
of polarized parton distributions in figs.3c and 3d are due to the differences 
in the $x$-shapes of the polarized parton distributions, particularly
those for $\Delta G$. All the asymmetries fall as $\eta^J$ goes to the 
extreme values as may be expected since the unpolarized gluon
distribution has a singular ($x^{-a}$ where $a < 1$) behaviour at small
$x$, whereas all the polarized gluon distributions go to zero at small
$x$. All the other differences between the asymmetries are due directly
to the differences between the corresponding polarized parton
distributions. This re-enforces the conclusion that this cross section
will undoubtedly yield very important information on the polarized
parton distributions.

\section{Conclusions}

We have examined the possibility that both the size and
$x$-shape of the polarized gluon distribution of the proton, $\Delta G$, may
be measured at RHIC via a measurement of the photon plus
jet cross section. Control over the kinematic variables of both the
photon and jet allows a much better determination of the $x$-value
probed when compared to inclusive prompt photon production. A comparison
of the predictions obtained using different polarized parton densities
show that a clear distinction between both the sizes and shapes should be
possible. Further more detailed information on the $x$-shapes of the
polarized gluon distribution will be obtainable by carefully choosing
the kinematic regions in which the jet and/or photon is tagged.

Assuming that the `fitted $\Delta G$' scenario is the most
plausible distribution, then a typical value for the asymmetry, $A_{LL}$
is $5\%$, but given the uncertainty in $\Delta G$ the asymmetry could
be as small as $1\%$ or as large as $20\%$. The expected small-$x$ behavior 
of the polarized and unpolarized distributions lead to differencs in  
correlations between the rapidities of the photon and jet in each case. 
This effect can produce very large asymmetries in certain regions of phase 
space which may be expolited and used to discriminate between various
models of $\Delta G$.

\section{Appendix}

The cross section for the 2-to-2 process is generically given by 

\beq
d\sigma={ |M|^2 (2 \pi)^4\over 4 p_1\cdot p_2}
\delta^4(p_1 + p_2 - k_1 - k_2) {\delta_+(k_1^2)\over (2\pi)^3}
{ \delta_+(k_2)^2\over (2 \pi)^3} \delta^4(p_1 + p_2 - k_1 -k_2),
\eeq
and working in the c.m frame 

\beq
{d\sigma\over d\, t}={1\over 16 \pi^2} |M|^2.
\eeq

We expand the final state momenta in the light cone variables 

\beqa
\sigma &\sim & \int d k_1^+ d k_1^- d^2k_{1 \perp} d k_2^+ d k_2^- 
d^2k_{2 \perp}\delta^{(2)}(k_{1\perp} + k_{2\perp})
\delta(p_1^+ + p_2^+ - k_1^+ - k_2^+)\nonumber \\
&& \,\,\, \times \delta(p_1^- + p_2^- - k_1^- - k_2^-)
\delta_+(k_1^2)\,\,\delta_+(k_2^2)\nonumber \\
\eeqa
and integrate over $k_1^-$, $k_2^+$ and $k_{2\perp}$, thereby eliminatigng 3 of the 5 delta functions. 

The remaining 2 delta functions are rewritten as 
\beqa
\delta(k_1^2)&=&\delta\left( \sqrt{2} k_\perp e^{y_1}(x_2 Q - 
k_\perp e^{- y_2}) - k_\perp^2 \right) \nonumber \\
\delta(k_2^2)&=&\delta\left( \sqrt{2} k_\perp e^{-y_2}(x_1 Q - 
k_\perp e^{y_1}) - k_\perp^2 \right) \nonumber \\
\eeqa

and the integration over the parton fractions $x_1,\,\, x_2$ performed by the relation 

\beq
\delta_+(k_1^2)\,\,\delta_+(k_2^2)={1\over 2 k_\perp^2 e^{\Delta y} Q^2}
\delta(x_1 - \overline{x}_1 )\delta(x_1 - \overline{x}_2)
\eeq

with 
\beqa
&& \overline{x}_1={k_\perp (1 + e^{\Delta y})\over \sqrt{2} Q e^{y_1}}\nonumber \\
&& \overline{x}_2={k_\perp (1 + e^{\Delta y})\over \sqrt{2} Q e^{-y_2}}.
\nonumber \\
\eeqa

Therefore, the convolution integral for the cross section is reorganized as follows

\beqa
d\sigma &\sim & \int_0^1 d x_1\int_0^1 d x_2 f(x_1) f(x_2) 
\int d k_1^+ d k_2^- d^2 k_\perp \delta(k_1^2) \delta(k_2^2) |M|^2
\nonumber \\
&=&  \int_0^1 d x_1\int_0^1 d x_2 f(x_1) f(x_2) 
\int d k_1^+ d k_2^- d^2 k_\perp  
{\delta(x_1 - \overline{x}_1 )\delta(x_1 - \overline{x}_2)\over 2 k_\perp^2 e^{\Delta y} Q^2} |M|^2 \nonumber \\
&=& f(\overline{x}_1) f(\overline{x}_2)\int {d k_1^+ d k_2^- d^2 k_\perp
\over 2 |k_\perp|^2 e^{\Delta y} Q^2}.
\eeqa 

At this point we change the remaining integration variables to the rapidity 
space using 
$\partial (k_1^+,k_2^-)/\partial(y_1,y_2)= exp[\Delta y]k_\perp^2/2$
and perform the integration over one of the angle in $\vec{k}_\perp$. 
With the correct normalization, we get the result in $\ref{mainq}$.

\section{Acknowledgments}

The work at Argonne National Laboratory was supported by the US Department of
Energy, Division of High Energy Physics, Contract number W-31-109-ENG-38.
\pagebreak


\pagebreak

\noindent
\begin{center}
{\large FIGURE CAPTIONS}
\end{center}
\newcounter{num}
\begin{list}%
{[\arabic{num}]}{\usecounter{num}
    \setlength{\rightmargin}{\leftmargin}}

\item (a) $p_T^\gamma$ distribution of the photon plus jet triple
differential cross section $d^3\sigma^{\gamma J}/dp_T^\gamma d
\eta^\gamma d\eta^J$ at $\sqrt{S}=500$ GeV for various polarized parton 
distributions and for rapidities of the photon and jet averaged over the region 
$-0.5\leq\eta^\gamma,\eta^J\leq 0.5$. The cut $p_T^J\geq 10$ GeV is
imposed.
(b) Similar to (a) at $\sqrt{S}=200$ GeV. (c) and (d) Corresponding
asymmetries for the distributions in (a) and (b) respectively.
\item (a) Distribution in the rapidity of the photon at $p_T^\gamma=10$
GeV and $\eta^J$ averaged over the region $-0.5\leq\eta^J\leq 0.5$ and 
$p_T^J\geq 10$ GeV at $\sqrt{S}=200 GeV$. (b) Similar to (a) but for 
$0.5\leq\eta^J\leq 1.5$. 
(c) and (d) Asymmetries for the curves shown in (a) and (b)
respectively.
\item (a) Distribution in the rapidity of the jet at $p_T^\gamma=10$
GeV and $\eta^\gamma$ averaged over the region $-0.5\leq\eta^\gamma\leq 0.5$ 
and $p_T^J\geq 10$ GeV at $\sqrt{S}=200 GeV$. (b) Similar to (a) but for 
$0.5\leq \eta^\gamma \leq 1.5$. 
(c) and (d) Asymmetries for the curves shown in (a) and (b)
respectively.
\end{list}
\pagebreak
\begin{figure}
\epsfxsize=155mm
\centerline{\epsfbox{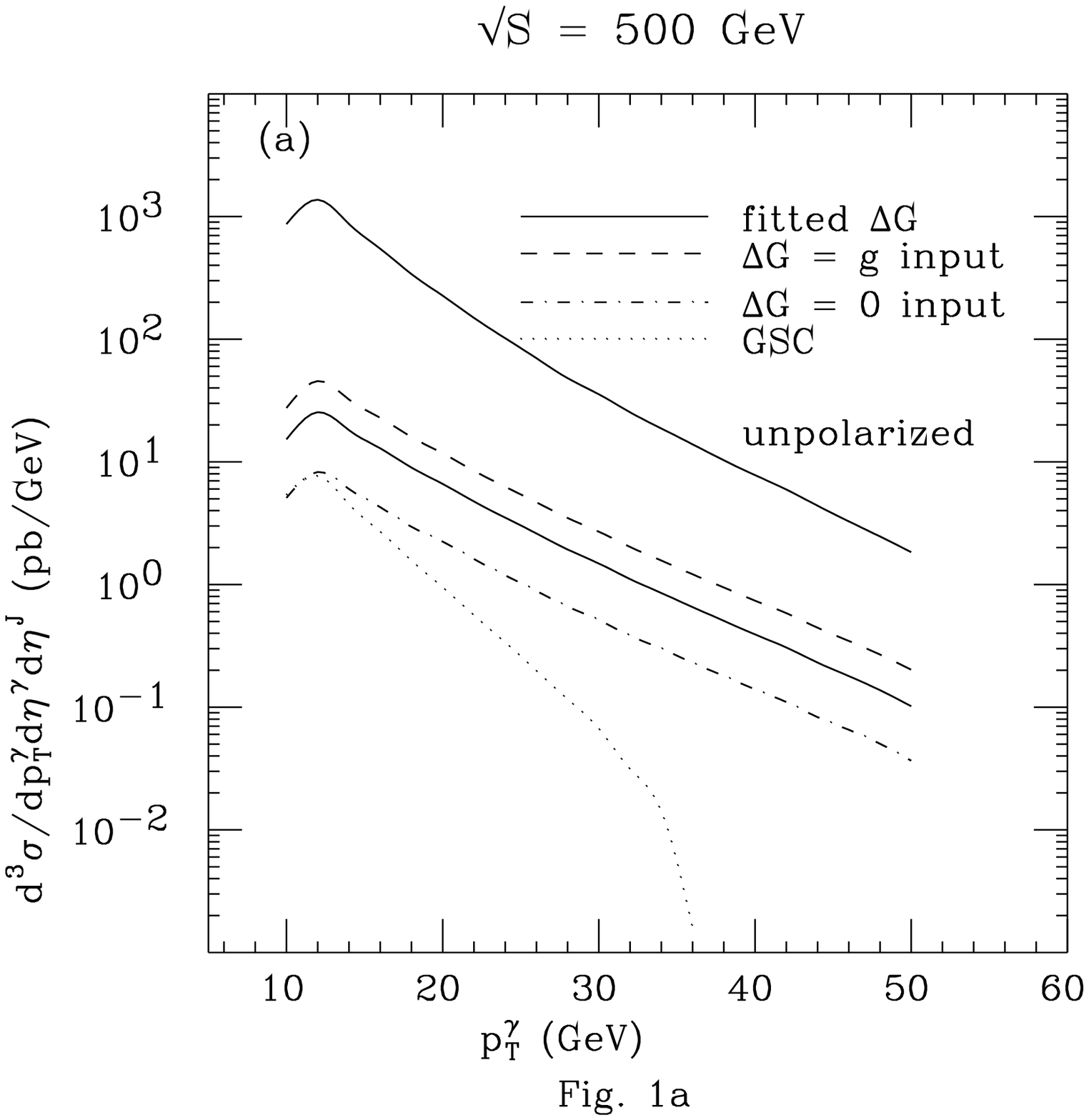}}
\epsfxsize=155mm
\centerline{\epsfbox{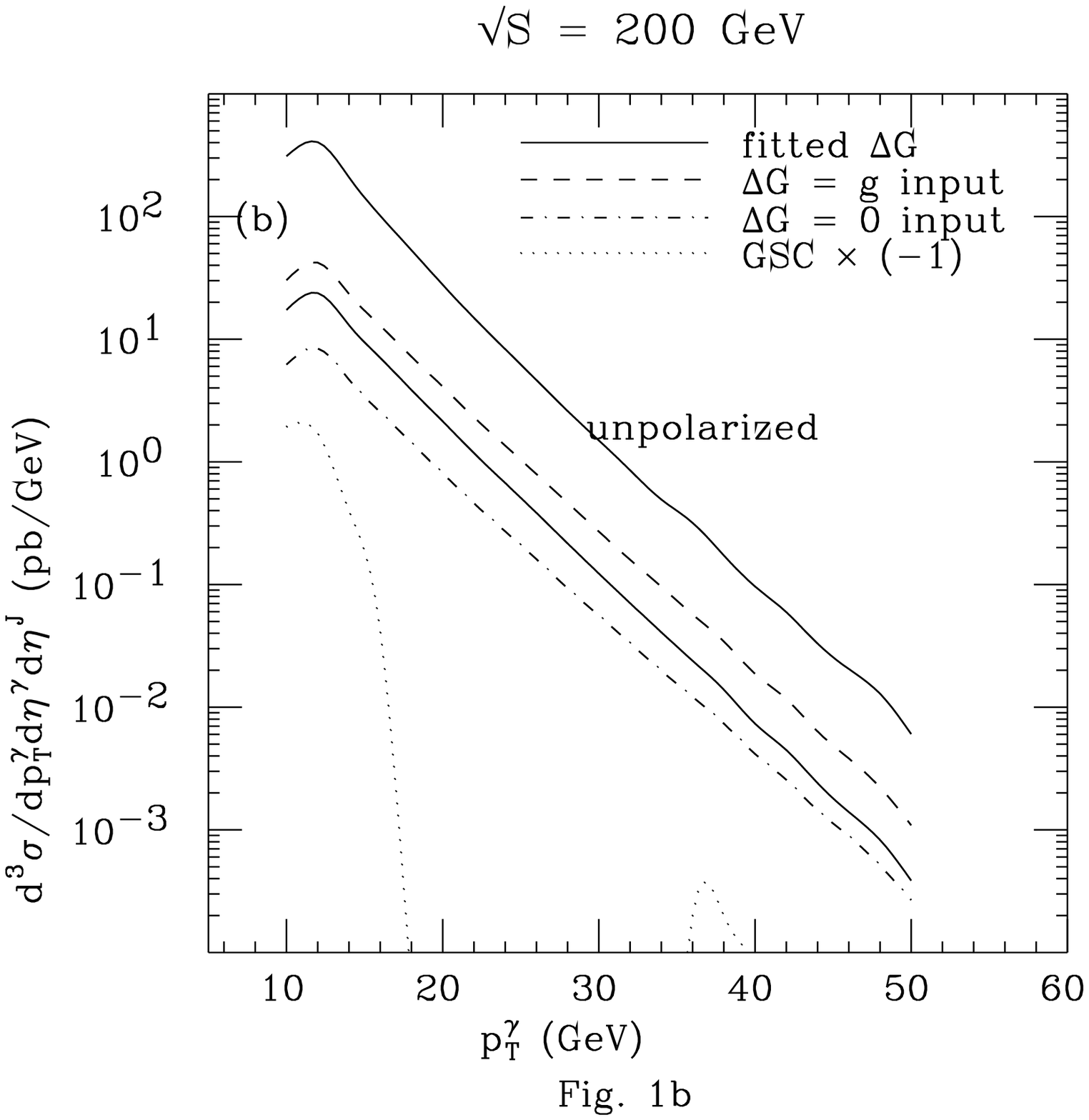}}
\epsfxsize=155mm
\centerline{\epsfbox{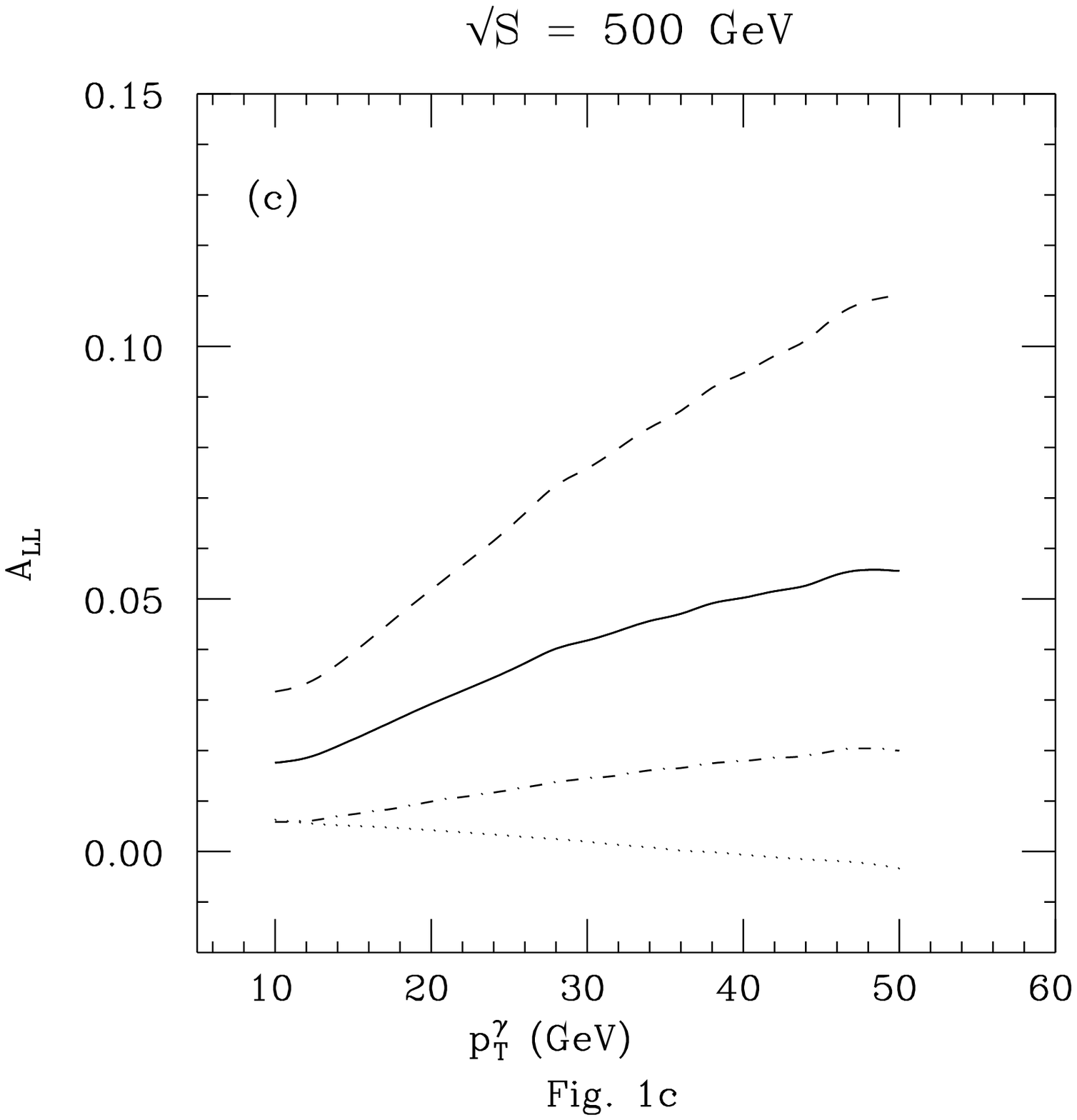}}
\epsfxsize=155mm
\centerline{\epsfbox{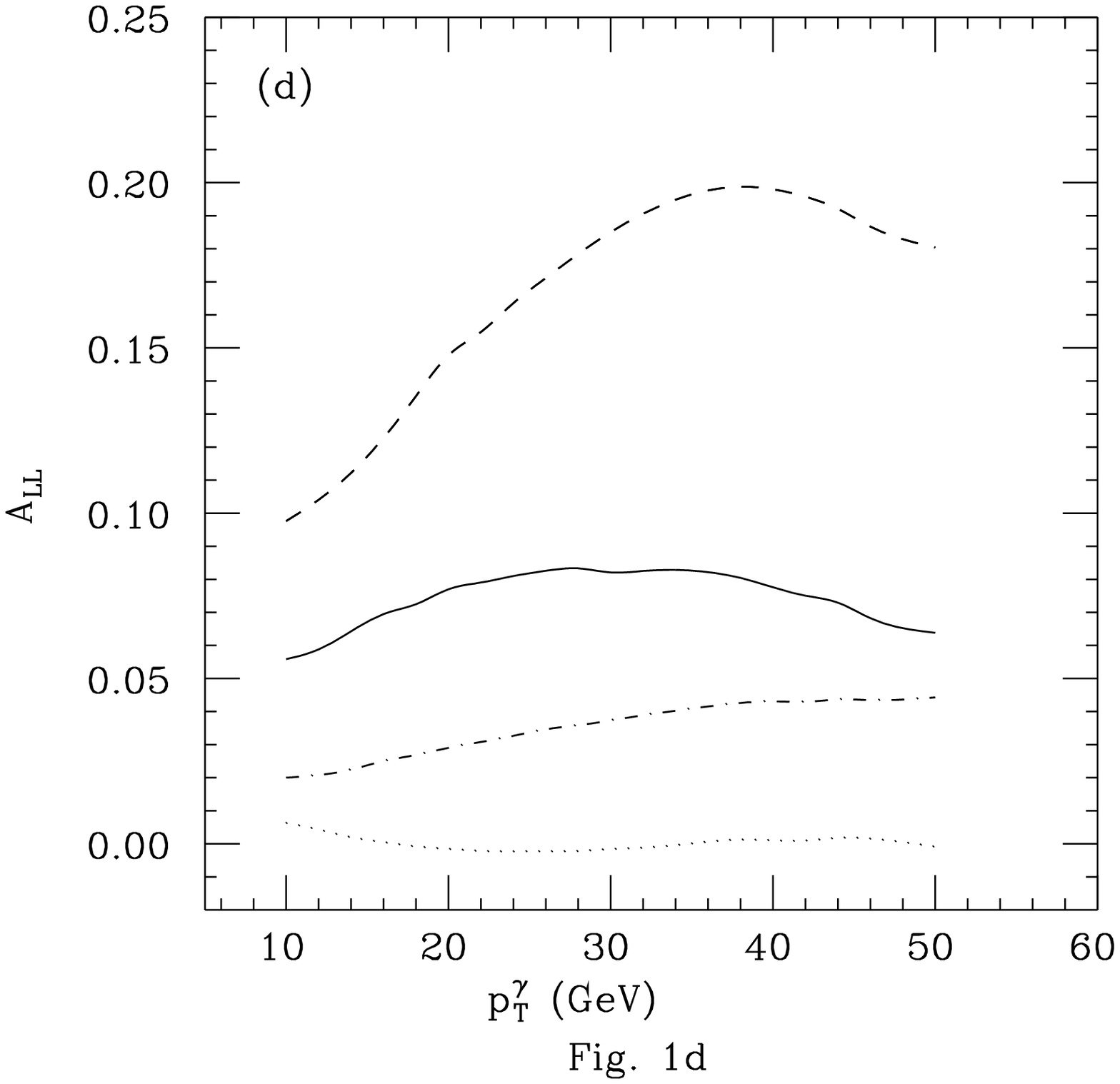}}
\epsfxsize=155mm
\centerline{\epsfbox{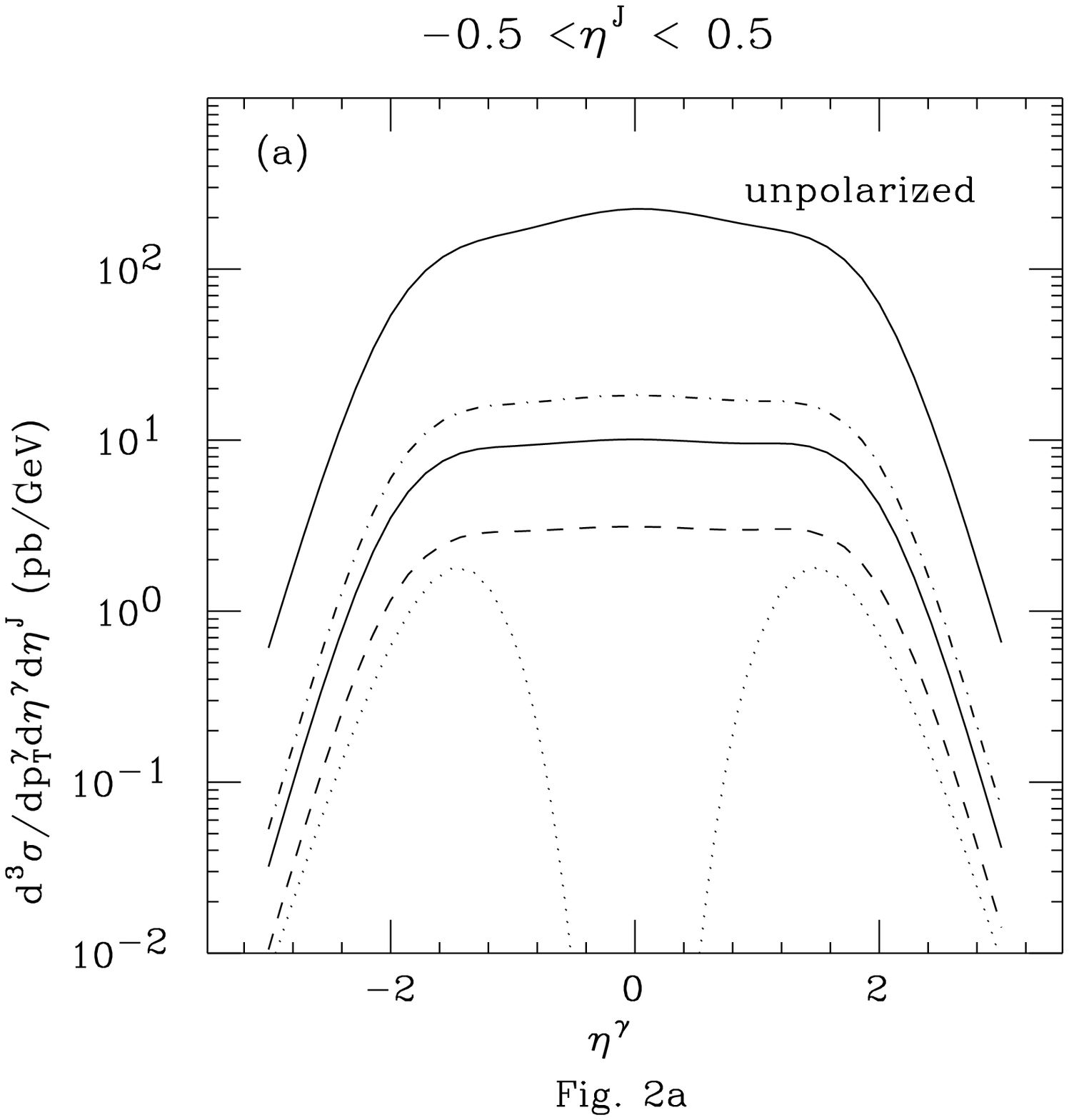}}
\epsfxsize=155mm
\centerline{\epsfbox{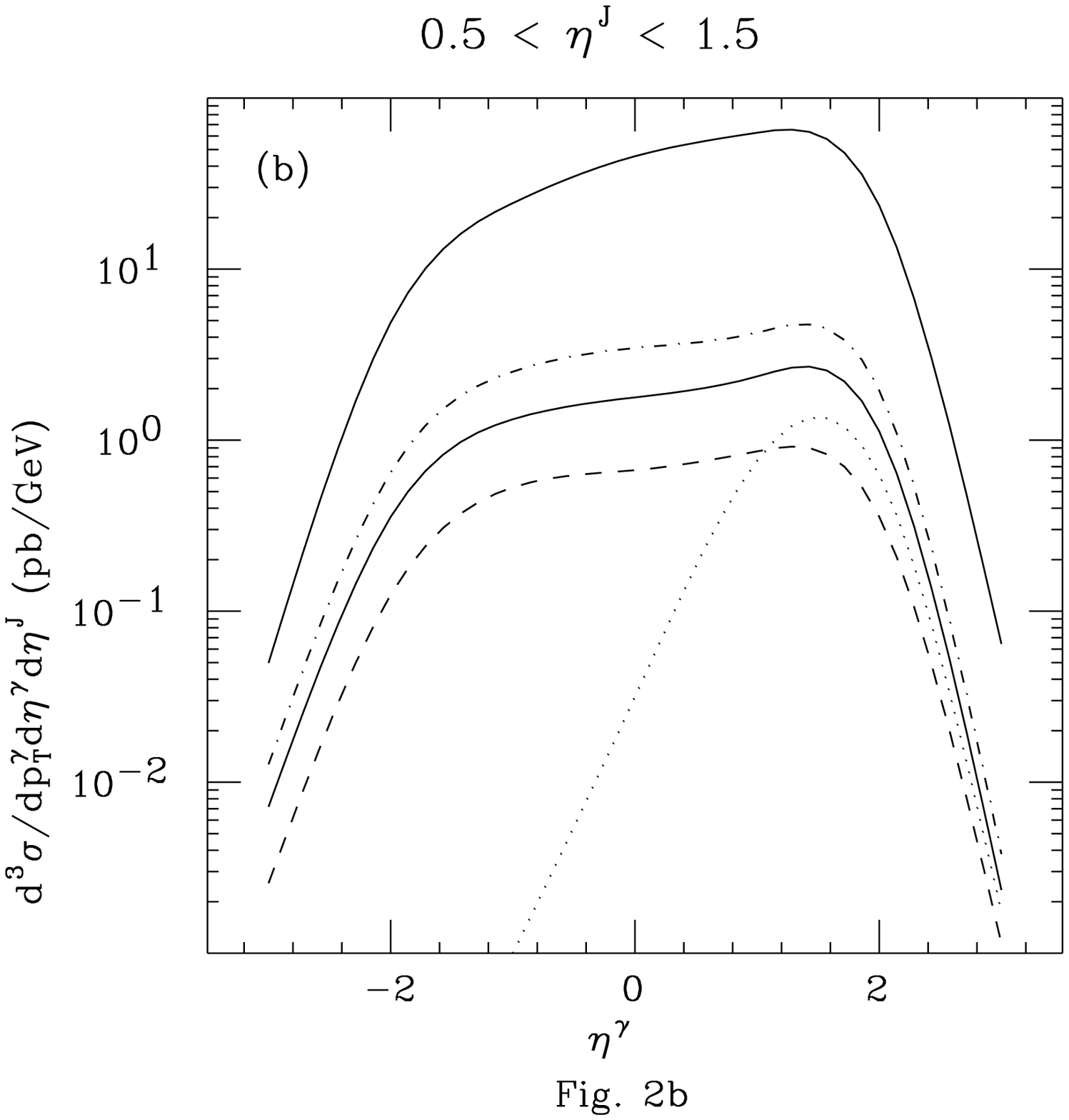}}
\epsfxsize=155mm
\centerline{\epsfbox{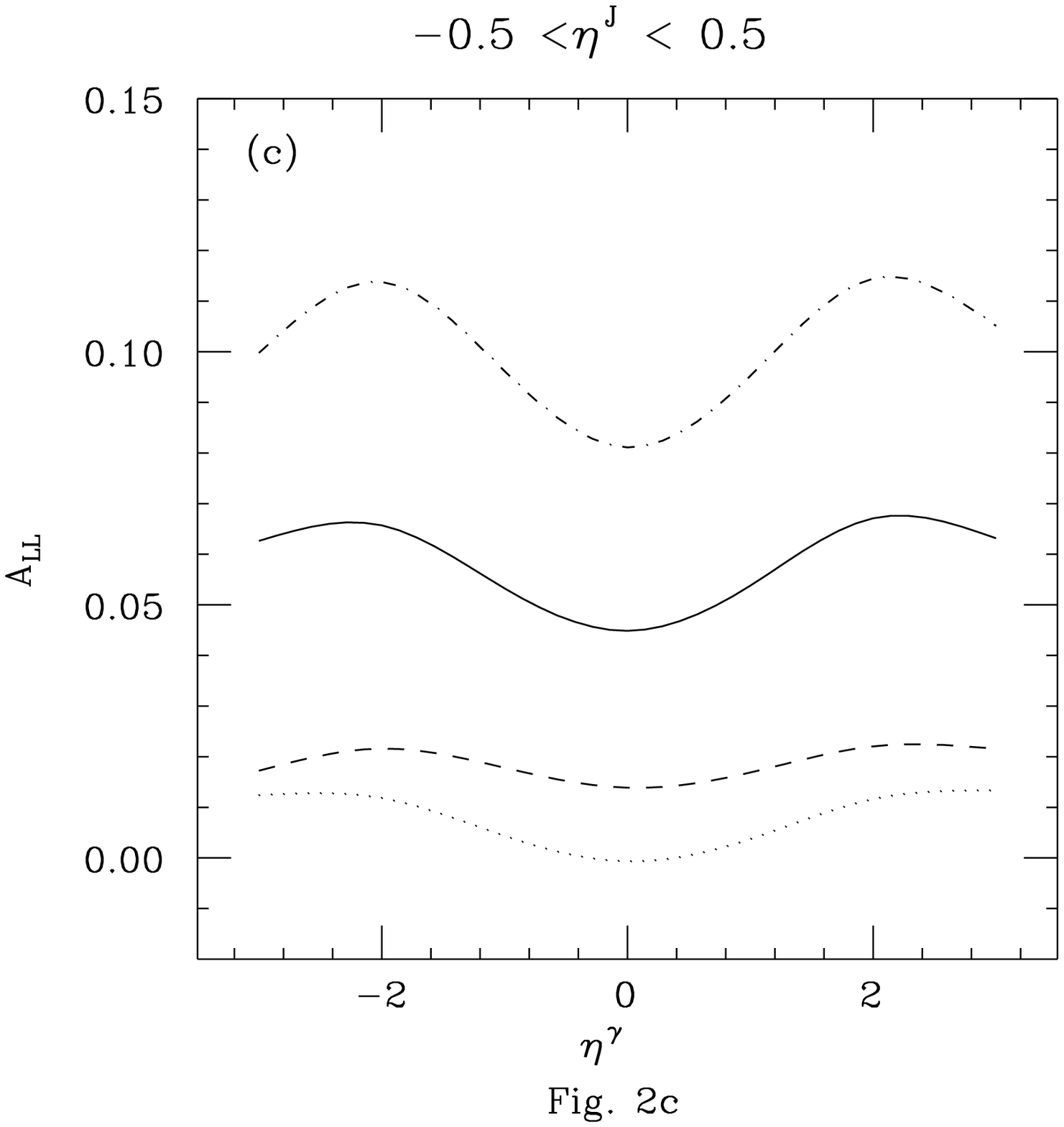}}
\epsfxsize=155mm
\centerline{\epsfbox{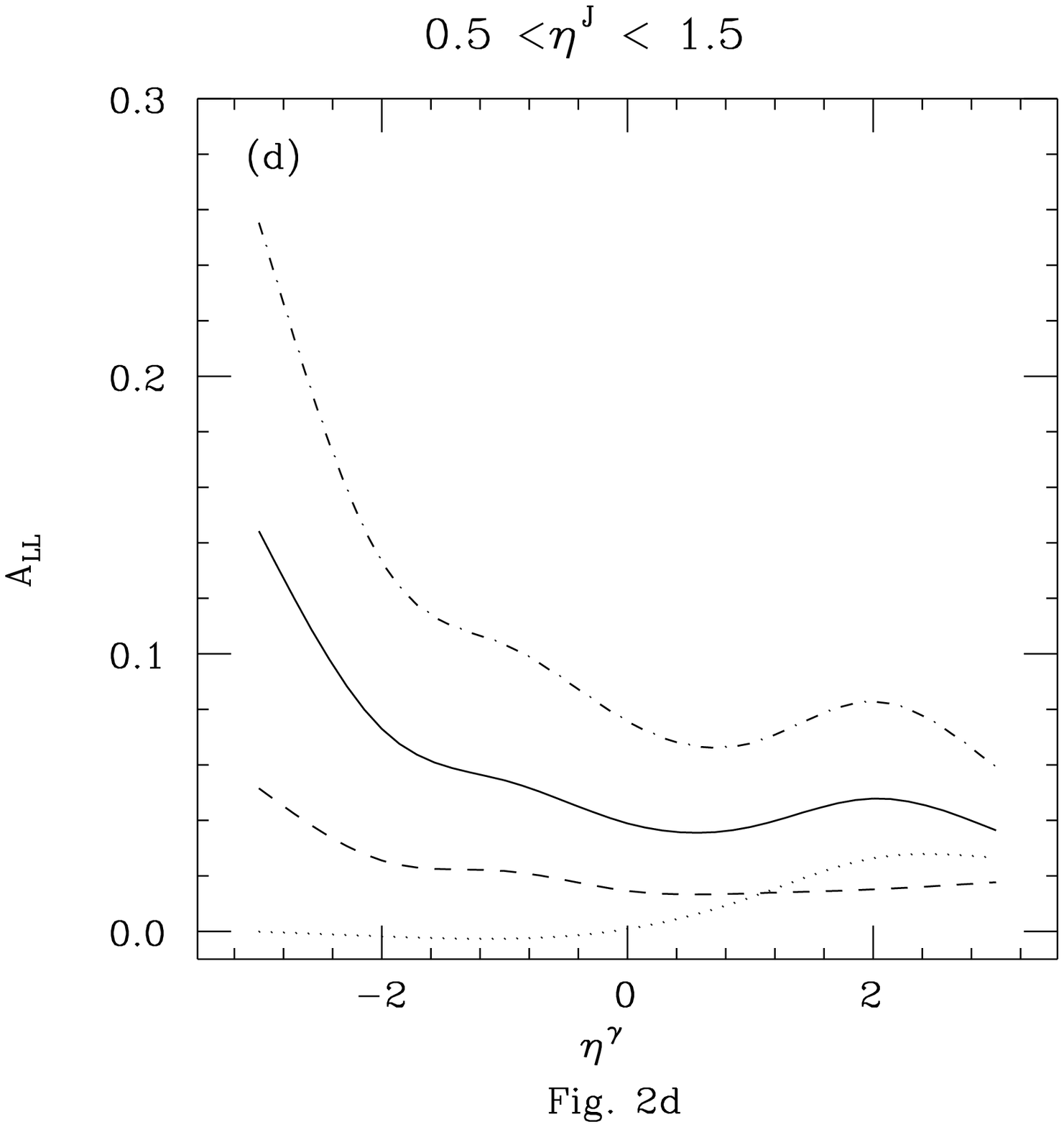}}
\epsfxsize=155mm
\centerline{\epsfbox{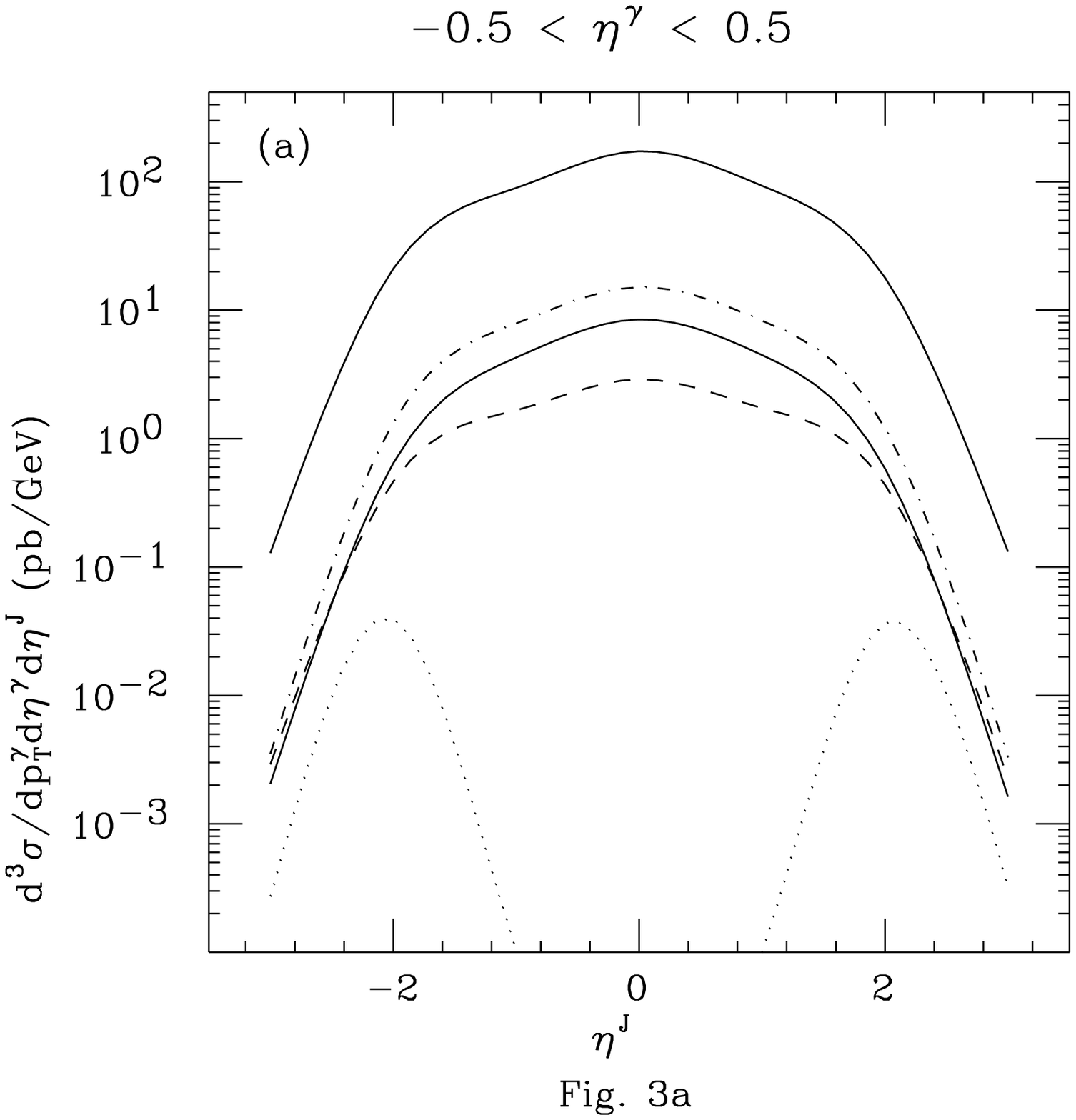}}
\epsfxsize=155mm
\centerline{\epsfbox{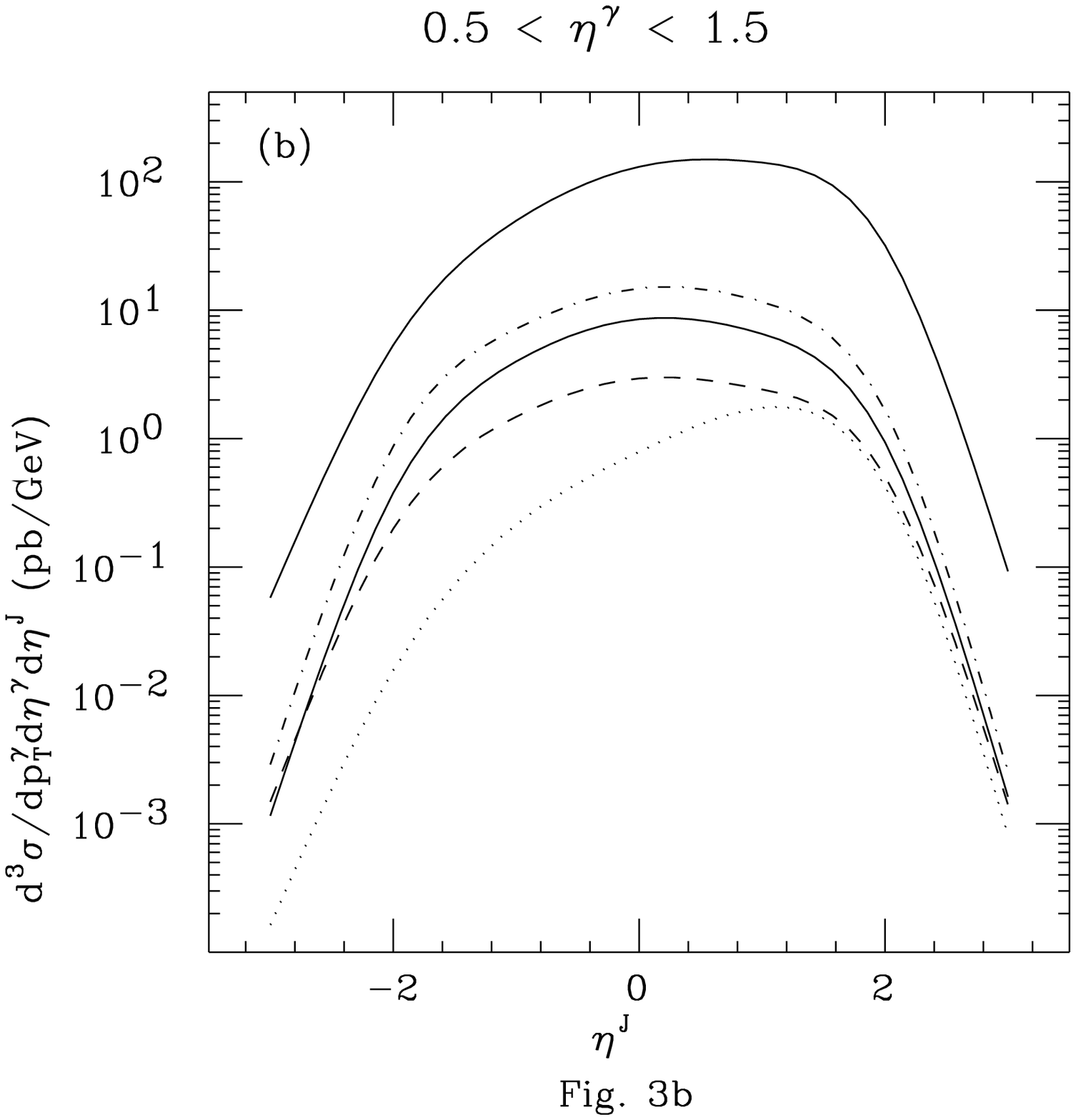}}
\epsfxsize=155mm
\centerline{\epsfbox{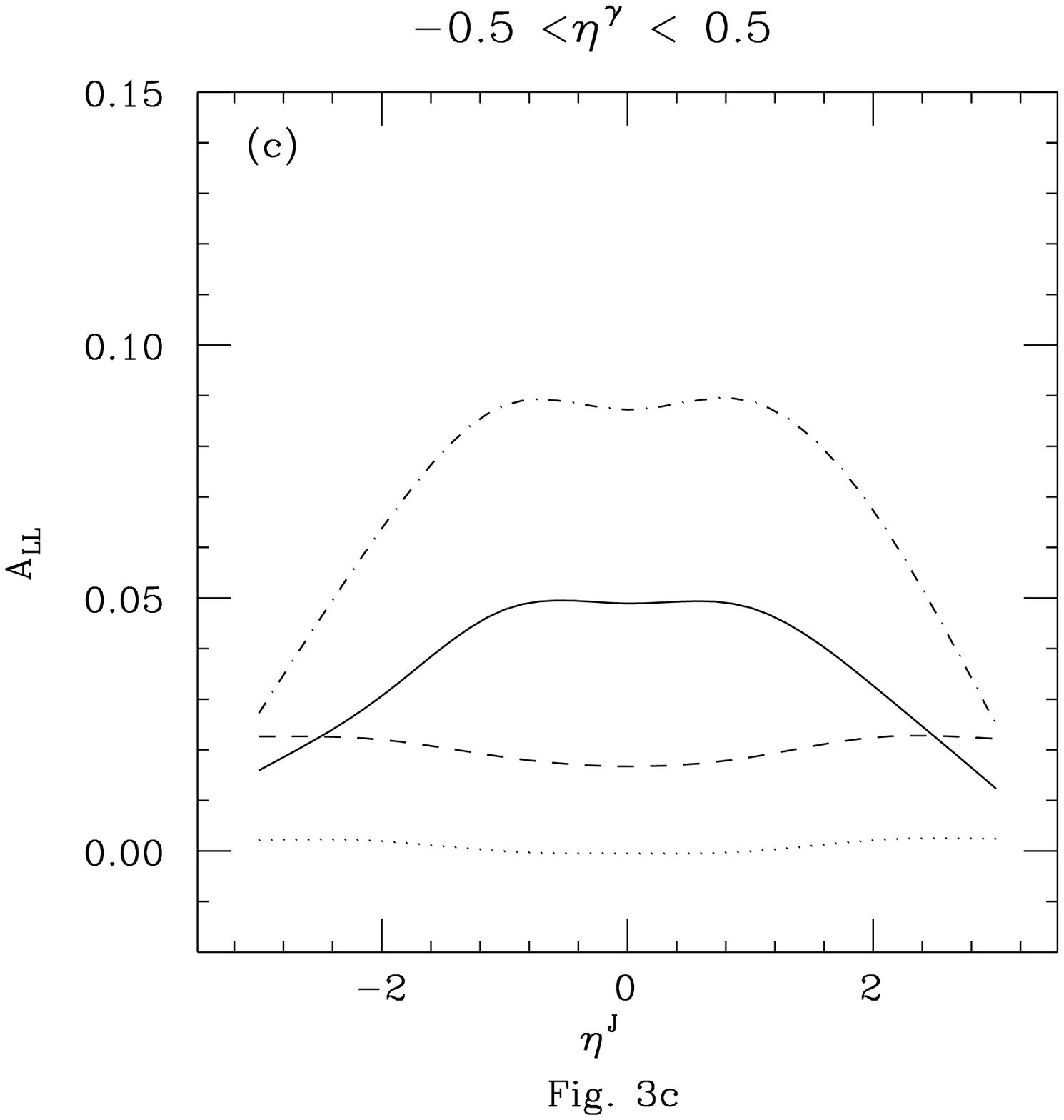}}
\epsfxsize=155mm
\centerline{\epsfbox{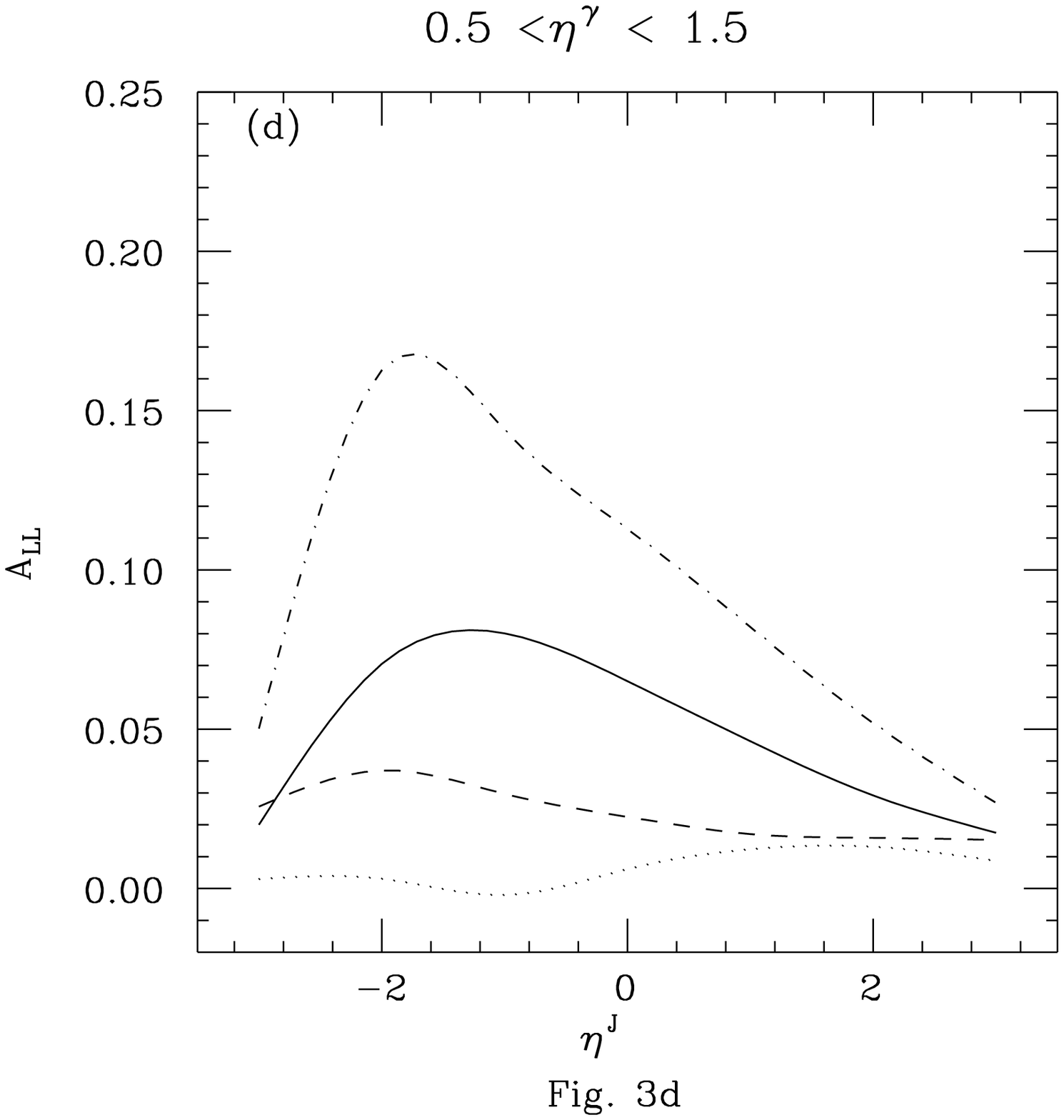}}
\setcounter{figure}{3}
\caption{}
\end{figure}
\end{document}